\documentclass[twocolumn,showpacs,amsmath,amssymb,superscriptaddress,nofootinbib]{revtex4-1}

\usepackage{graphicx}
\usepackage{dcolumn}
\usepackage{bm}
\usepackage{color}
\usepackage{ulem}
\usepackage{url}
\usepackage{multirow}

\newcommand{\bfg }{\begin{figure}[htpb]}
\newcommand{\efg }{\end{figure}}
\newcommand{\bmn }{\begin{minipage}}
\newcommand{\emn }{\end{minipage}}
\newcommand{\bt }{\begin{table}[htpb]}
\newcommand{\et }{\end{table}}

\newcommand{\sNN}{$\sqrt{s_{\rm NN}}$ }

\newcommand{\GeVcsq}{GeV/$c^2$ }

\newcommand{ \be }{\begin{equation}}
\newcommand{ \ee }{\end{equation}}
\newcommand{ \bea }{\begin{eqnarray}}
\newcommand{ \eea }{\end{eqnarray}}

\newcommand{ \pT }{$p_{\rm T}$ }

\begin{document}
\def\Journal#1#2#3#4{{#1} {\bf #2}, #3 (#4)}

\def\NCA{Nuovo Cimento}
\def\NIM{Nucl. Instr. Meth.}
\def\NIMA{{Nucl. Instr. Meth.} A}
\def\NPB{{Nucl. Phys.} B}
\def\NPA{{Nucl. Phys.} A}
\def\PLB{{Phys. Lett.}  B}
\def\PRL{{Phys. Rev. Lett.}}
\def\PRC{{Phys. Rev.} C}
\def\PRD{{Phys. Rev.} D}
\def\ZPC{{Z. Phys.} C}
\def\JPG{{J. Phys.} G}
\def\EPJ{{Eur. Phys. J.} C}
\def\EPJA{{Eur. Phys. J.} A}
\def\EPJST{{Eur. Phys. J.} - Special Topics}
\def\RPP{{Rep. Prog. Phys.}}
\def\PR{{Phys. Rep.}}
\def\ANP{{Adv. Nucl. Phys.}}



\title{Dielectron Mass Spectra from Au+Au Collisions at \sNN = 200\,GeV}

\affiliation{AGH University of Science and Technology, Cracow, Poland}
\affiliation{Argonne National Laboratory, Argonne, Illinois 60439, USA}
\affiliation{University of Birmingham, Birmingham, United Kingdom}
\affiliation{Brookhaven National Laboratory, Upton, New York 11973, USA}
\affiliation{University of California, Berkeley, California 94720, USA}
\affiliation{University of California, Davis, California 95616, USA}
\affiliation{University of California, Los Angeles, California 90095, USA}
\affiliation{Universidade Estadual de Campinas, Sao Paulo, Brazil}
\affiliation{Central China Normal University (HZNU), Wuhan 430079, China}
\affiliation{University of Illinois at Chicago, Chicago, Illinois 60607, USA}
\affiliation{Cracow University of Technology, Cracow, Poland}
\affiliation{Creighton University, Omaha, Nebraska 68178, USA}
\affiliation{Czech Technical University in Prague, FNSPE, Prague, 115 19, Czech Republic}
\affiliation{Nuclear Physics Institute AS CR, 250 68 \v{R}e\v{z}/Prague, Czech Republic}
\affiliation{Frankfurt Institute for Advanced Studies FIAS, Germany}
\affiliation{Institute of Physics, Bhubaneswar 751005, India}
\affiliation{Indian Institute of Technology, Mumbai, India}
\affiliation{Indiana University, Bloomington, Indiana 47408, USA}
\affiliation{Alikhanov Institute for Theoretical and Experimental Physics, Moscow, Russia}
\affiliation{University of Jammu, Jammu 180001, India}
\affiliation{Joint Institute for Nuclear Research, Dubna, 141 980, Russia}
\affiliation{Kent State University, Kent, Ohio 44242, USA}
\affiliation{University of Kentucky, Lexington, Kentucky, 40506-0055, USA}
\affiliation{Korea Institute of Science and Technology Information, Daejeon, Korea}
\affiliation{Institute of Modern Physics, Lanzhou, China}
\affiliation{Lawrence Berkeley National Laboratory, Berkeley, California 94720, USA}
\affiliation{Massachusetts Institute of Technology, Cambridge, MA 02139-4307, USA}
\affiliation{Max-Planck-Institut f\"ur Physik, Munich, Germany}
\affiliation{Michigan State University, East Lansing, Michigan 48824, USA}
\affiliation{Moscow Engineering Physics Institute, Moscow Russia}
\affiliation{National Institute of Science Education and Research, Bhubaneswar 751005, India}
\affiliation{Ohio State University, Columbus, Ohio 43210, USA}
\affiliation{Old Dominion University, Norfolk, VA, 23529, USA}
\affiliation{Institute of Nuclear Physics PAN, Cracow, Poland}
\affiliation{Panjab University, Chandigarh 160014, India}
\affiliation{Pennsylvania State University, University Park, Pennsylvania 16802, USA}
\affiliation{Institute of High Energy Physics, Protvino, Russia}
\affiliation{Purdue University, West Lafayette, Indiana 47907, USA}
\affiliation{Pusan National University, Pusan, Republic of Korea}
\affiliation{University of Rajasthan, Jaipur 302004, India}
\affiliation{Rice University, Houston, Texas 77251, USA}
\affiliation{Universidade de Sao Paulo, Sao Paulo, Brazil}
\affiliation{University of Science \& Technology of China, Hefei 230026, China}
\affiliation{Shandong University, Jinan, Shandong 250100, China}
\affiliation{Shanghai Institute of Applied Physics, Shanghai 201800, China}
\affiliation{SUBATECH, Nantes, France}
\affiliation{Temple University, Philadelphia, Pennsylvania, 19122, USA}
\affiliation{Texas A\&M University, College Station, Texas 77843, USA}
\affiliation{University of Texas, Austin, Texas 78712, USA}
\affiliation{University of Houston, Houston, TX, 77204, USA}
\affiliation{Tsinghua University, Beijing 100084, China}
\affiliation{United States Naval Academy, Annapolis, MD 21402, USA}
\affiliation{Valparaiso University, Valparaiso, Indiana 46383, USA}
\affiliation{Variable Energy Cyclotron Centre, Kolkata 700064, India}
\affiliation{Warsaw University of Technology, Warsaw, Poland}
\affiliation{University of Washington, Seattle, Washington 98195, USA}
\affiliation{Yale University, New Haven, Connecticut 06520, USA}
\affiliation{University of Zagreb, Zagreb, HR-10002, Croatia}

\author{L.~Adamczyk}\affiliation{AGH University of Science and Technology, Cracow, Poland}
\author{J.~K.~Adkins}\affiliation{University of Kentucky, Lexington, Kentucky, 40506-0055, USA}
\author{G.~Agakishiev}\affiliation{Joint Institute for Nuclear Research, Dubna, 141 980, Russia}
\author{M.~M.~Aggarwal}\affiliation{Panjab University, Chandigarh 160014, India}
\author{Z.~Ahammed}\affiliation{Variable Energy Cyclotron Centre, Kolkata 700064, India}
\author{I.~Alekseev}\affiliation{Alikhanov Institute for Theoretical and Experimental Physics, Moscow, Russia}
\author{J.~Alford}\affiliation{Kent State University, Kent, Ohio 44242, USA}
\author{C.~D.~Anson}\affiliation{Ohio State University, Columbus, Ohio 43210, USA}
\author{A.~Aparin}\affiliation{Joint Institute for Nuclear Research, Dubna, 141 980, Russia}
\author{D.~Arkhipkin}\affiliation{Brookhaven National Laboratory, Upton, New York 11973, USA}
% ...

\collaboration{STAR Collaboration}\noaffiliation

\date{\today}

\begin{abstract}
We report the STAR measurements  of dielectron ($e^+e^-$) production at midrapidity ($|y_{ee}|<$1) in Au+Au collisions at \sNN = 200\,GeV. The measurements are evaluated in different invariant mass regions with a focus on 0.30-0.76 ($\rho$-like), 0.76-0.80 ($\omega$-like), and 0.98-1.05 ($\phi$-like)\,GeV/$c^{2}$. The spectrum in the $\omega$-like and $\phi$-like regions can be well described by the hadronic cocktail simulation. In the $\rho$-like region, however, the vacuum $\rho$ spectral function cannot describe the shape of the dielectron excess. In this range, an enhancement of 1.77$\pm$0.11(stat.)$\pm$0.24(sys.)$\pm$0.33(cocktail) is determined with respect to the hadronic cocktail simulation that excludes the $\rho$ meson.
The excess yield in the $\rho$-like region increases with the number of collision participants faster than the $\omega$ and $\phi$ yields. Theoretical models with broadened $\rho$ contributions through interactions with constituents in the hot QCD medium provide a consistent description of the dilepton mass spectra for the measurement presented here and the earlier data at the Super Proton Synchrotron energies.

\end{abstract}
\maketitle


Recent results from Relativistic Heavy Ion Collider (RHIC) continue to provide mounting evidence that a strongly coupled Quark-Gluon Plasma (QGP) has been created in the Au+Au collisions at \sNN = 200\,GeV~\cite{STARwhitepaper}.
One of the scientific goals of the current high energy heavy ion program is to quantify properties of this QGP matter, such as the equation of state and the intrinsic chiral characteristics.
Dileptons are a clean and penetrating probe for studying these properties because leptons do not suffer from strong interactions.

Thermal dileptons radiated from the partonic medium have been suggested as a unique probe for temperature measurement of the QGP~\cite{Shuryak}.
Theoretical calculations suggest that at RHIC energies, QGP thermal dilepton production becomes significant at dilepton invariant mass $M_{ll}>$1\,GeV/$c^2$, with increasingly higher masses corresponding to earlier stages of the production~\cite{RappThermalEE}.
As the system cools, dileptons emitted from the hadronic medium are governed by the coupling of vector mesons (e.g., $\rho$) to the medium and are expected to dominate the low-mass production ($M_{ll}<$1\,GeV/$c^2$)~\cite{RappWambach}. Their vacuum mass spectra are determined by the spontaneously broken chiral symmetry. Theoretical calculations, however, suggest that vector meson spectra will be modified in a hot and dense medium, reflecting the restoration of the broken chiral symmetry~\cite{BrownRho,RappWambach2}.
After freeze-out, long lived particles ($\pi^0$, $\eta$, $D\overline{D}$, etc.) can decay to lepton pairs. The sum of these contributions, usually referred as a hadronic cocktail, can be calculated based on the measured or estimated yields.

Dilepton measurements have been a subject of experimental investigations since the early days of heavy ion collisions~\cite{DLS,HADES,HELIOS3,CERES,NA60,PHENIX}.
Of particular interest, measurements from CERES and NA60 (\sNN = 8.75\,-\,17.2\,GeV)
showed a clear enhancement in the mass region below $\sim$0.7\,\GeVcsq when compared to known hadronic sources. High precision data from NA60 demonstrated that the enhancement is consistent with in-medium broadening of the $\rho$ mass spectrum instead of a dropping mass hypothesis ~\cite{NA60,Rapp4SPS,Renk4SPS,Dusling4SPS,PHSD4SPS}. NA60 also observed that after removing correlated charm contributions, slope parameters of dimuon transverse mass spectra show a sudden change above the $\phi$ mass, which is argued to be indicative of the partonic thermal dileptons presence~\cite{NA60}.
At RHIC, PHENIX reported a significant enhancement in the mass region of 0.30-0.76\,GeV/$c^2$ in Au+Au collisions at \sNN = 200\,GeV. The enhancement was predominantly at low transverse momentum ($p_{\rm T}$) and for the most central collisions~\cite{PHENIX}.

Experimental measurements suggest that the QCD medium at top RHIC energies undergoes a much longer partonic phase than at SPS energies~\cite{STARwhitepaper}. Moreover, the typical net baryon densities are found to be significantly different between the two energy regimes~\cite{PiKPSpectra}. Nevertheless, model calculations that successfully described the Super Proton Synchrotron (SPS) data~\cite{Rapp4SPS,Renk4SPS,Dusling4SPS,PHSD4SPS} expect the low-mass dilepton production to remain dominated by the vector meson contributions from the hadronic phase  with its spectral function broadening governed by the total baryon density~\cite{RappThermalEE}.
Therefore, dielectron measurements at RHIC energies can provide a clear probe into the production mechanisms as well as the evolution dynamics of these systems.
The magnitude of the dielectron excess reported by the PHENIX, however, is yet to be reproduced by such models.

In this Letter, we report the measurement at STAR of dielectron production in Au+Au collisions at \sNN = 200\,GeV. Data used in this analysis were recorded in the 2010 RHIC run, which includes 2.4$\times 10^8$ minimum bias (0-80\%) and 2.2$\times 10^8$ central (0-10\%) Au+Au events. The main subsystems used for the analysis are the Time Projection Chamber (TPC)~\cite{tpcnim} and the Time-Of-Flight (TOF) detectors~\cite{tofnim}.

Electron candidates (including positrons if not specified) were reconstructed in the TPC and required to have more than 20 out of a maximally possible 45 track-fit points to ensure sufficient momentum resolution. Each candidate should have at least 16 hit points that can be used for the determination of the specific energy-loss ($dE/dx$) in the TPC gas.
Electron candidates were also required to originate from the collision vertex based on an extrapolated distance of closest approach (DCA) to this vertex of less than 1 cm.
Electrons were identified via a combination of the $dE/dx$ measurement and the velocity measurement from the TOF~\cite{PIDNIM}.
The electron sample purity integrated over the measured $p_{\rm T}$ region was ensured to be at least 95\%
in order to keep correlated residuals due to hadron contamination to be less than 10\% of the signal.
All electron candidates with $p_{\rm T}>$0.2\,GeV/$c$ and pseudo-rapidity $|\eta|<$1 from the same event were combined to generate the unlike-sign pair distribution at midrapidity ($|y_{ee}|<$1). Dielectron pairs from photon conversion in materials were suppressed by the DCA selection and further reduced by a cut on the minimum pair opening angle~\cite{PHENIX}.

We adopted two approaches to reproduce the background that do not originate from pair production: the like-sign pair combinations and the mixed-event technique, for which unlike-sign pairs from different events were used.
In the low-mass region, like-sign pairs better reproduce the background spectrum, compared to mixed-event techniques, because the unlike-sign background contains residual correlations (e.g., conversion of photon pairs)~\cite{STARpp}. For this reason, the same-event like-sign distribution, corrected for the acceptance differences between like-sign and unlike-sign pairs, was used as the background for $M_{ll}<$1\,GeV/$c^2$.
At pair masses above 1\,\GeVcsq where the statistics of the like-sign distribution become limited, mixed-event distributions were used to evaluate the background. The mixed-event unlike-sign distribution, after
normalization, provides a good description of the uncorrelated combinatorial background. The normalization factor was determined based on the same-event like-sign and the mixed-event like-sign distributions in the mass region of 1-2\,GeV/$c^2$.
The normalized mixed-event unlike-sign distribution agrees with the same-event like-sign distribution above 1\,GeV/$c^{2}$ within uncertainties, but the centroid value falls slightly below the like-sign trend with increasing mass, which is attributed to residual correlated background (e.g., jet fragments).
The total background at $M_{ll}>$1\,GeV/$c^2$ therefore includes the combinatorial background using the mixed-event unlike-sign pairs and the residual correlated background based on a parametrization to the data.

\begin{figure}[ht]
\centering
\includegraphics[width=0.45\textwidth]{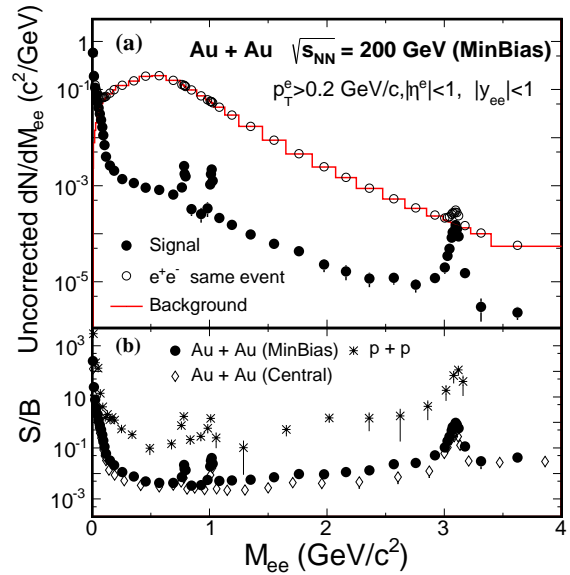}
\caption[]{(Color online) (a) Uncorrected distributions of $e^+e^-$ invariant mass (open circles), the reconstructed background (red histogram) and the signal (solid dots) pairs in 200 GeV Au+Au minimum bias collisions. (b) The ratio of signal to background in $p$+$p$~\cite{STARpp} and Au+Au collisions at \sNN = 200 GeV.}
\label{bg1}
\end{figure}

The raw distributions of $e^+e^-$ invariant mass, the reconstructed background and the background-subtracted signal in 200 GeV Au+Au minimum bias collisions are shown in Fig.~\ref{bg1} (a). The signal-to-background ratios from $p$+$p$~\cite{STARpp}, Au+Au minimum bias and Au+Au central collisions at \sNN = 200 GeV are shown in Fig.~\ref{bg1} (b).


The raw signal was corrected for the detector tracking efficiency.
The single electron tracking efficiency was determined using the embedding technique in which Monte Carlo (MC) simulated electrons, propagated through the STAR detector in GEANT~\cite{geant:321} to produce raw signals, are embedded into real data prior to processing with the offline reconstruction software.
The TOF matching and particle identification efficiencies were evaluated based on measured distributions from the real data~\cite{Efficiency}. The dielectron pair efficiency was obtained by convoluting with single electron efficiency and its kinematic dependence. Finally, we corrected for an additional inefficiency at very low-mass due to the photon conversion pair cut.

The systematic uncertainty in our final mass spectrum is estimated from the following sources: a$)$ uncertainty on the normalization factor for mixed-event background subtraction, 0.06\%, resulting in up to a maximum of 15\% uncertainty at 1\,GeV/$c^2$; b$)$ uncertainty on the residual correlated background resulting in 10\% uncertainty in the mass region of 1-3\,GeV/$c^2$ ; c$)$ uncertainty on the correction factor of the acceptance difference between like-sign and unlike-sign pairs resulting in 5-8\% uncertainty at 0.3-1\,GeV/$c^2$; d$)$ uncertainty due to the hadron contamination in the electron samples resulting in 8\% uncertainty between 1-3\,GeV/$c^2$ and e$)$ uncertainty on the detector efficiency correction, 13\%. The systematic uncertainty in the raw dielectron spectrum is evaluated as the sum of components a$)$-d$)$ applied at each pair mass region. The final systematic uncertainty is calculated as the quadratic sum of the raw spectrum uncertainty and the pair detection efficiency uncertainty e$)$.

In order to disentangle the various sources contributing to the dielectron signal, a hadronic cocktail simulation, performed previously for $p$+$p$ collisions~\cite{STARpp}, was generated for Au+Au collisions at \sNN = 200\,GeV. The simulation included dielectron contributions from decays or Dalitz decays of $\pi^{0}$, $\eta$, $\eta^{\prime}$, $\omega$, $\phi$, $J/\psi$, $c\bar{c}$, $b\bar{b}$ and Drell-Yan (DY) production.
Tsallis blast-wave parametrizations~\cite{TBW} based on RHIC measurements of light hadron spectra ($\pi^{\pm}$, $K^{\pm}$, $\phi$, $p$, $\bar{p}$, high-$p_{\rm T}$ $\eta$)~\cite{DataPion,DataEta,DataPhi,DataJpsi} were used as the inputs to our cocktail simulations.
The same parameters were also applied to mesons which have not yet been measured ($\omega$, $\eta$ at low-$p_{\rm T}$, and $\eta^{\prime}$). The dielectron yields from correlated charm or bottom decays and DY production are based on PYTHIA model~\cite{pythia} calculations in which parameters have been tuned to published STAR measurements~\cite{DataCharm}.
The input charm pair production cross section is $d\sigma/dy|_{y=0}$ = 171 $\pm$ 26 $\mu$b per nucleon-nucleon collision~\cite{DataCharm} and the $e^+e^-$ pairs from correlated charm decays in PYTHIA were scaled by the number-of-binary-collisions ($N_{\rm bin}$) to obtain the contribution in the Au+Au collisions.
The systematic uncertainty on the cocktail is dominated by the experimental uncertainties on the measured particle yields and spectra. In particular, the large uncertainty in the cocktail in the mass region of 0.15-1\,GeV/$c^2$ is mainly attributed to the unmeasured low-$p_{\rm T}$ $\eta$ mesons and the input charm cross section.

\begin{figure}[h]
\centering
\includegraphics[width=0.45\textwidth]{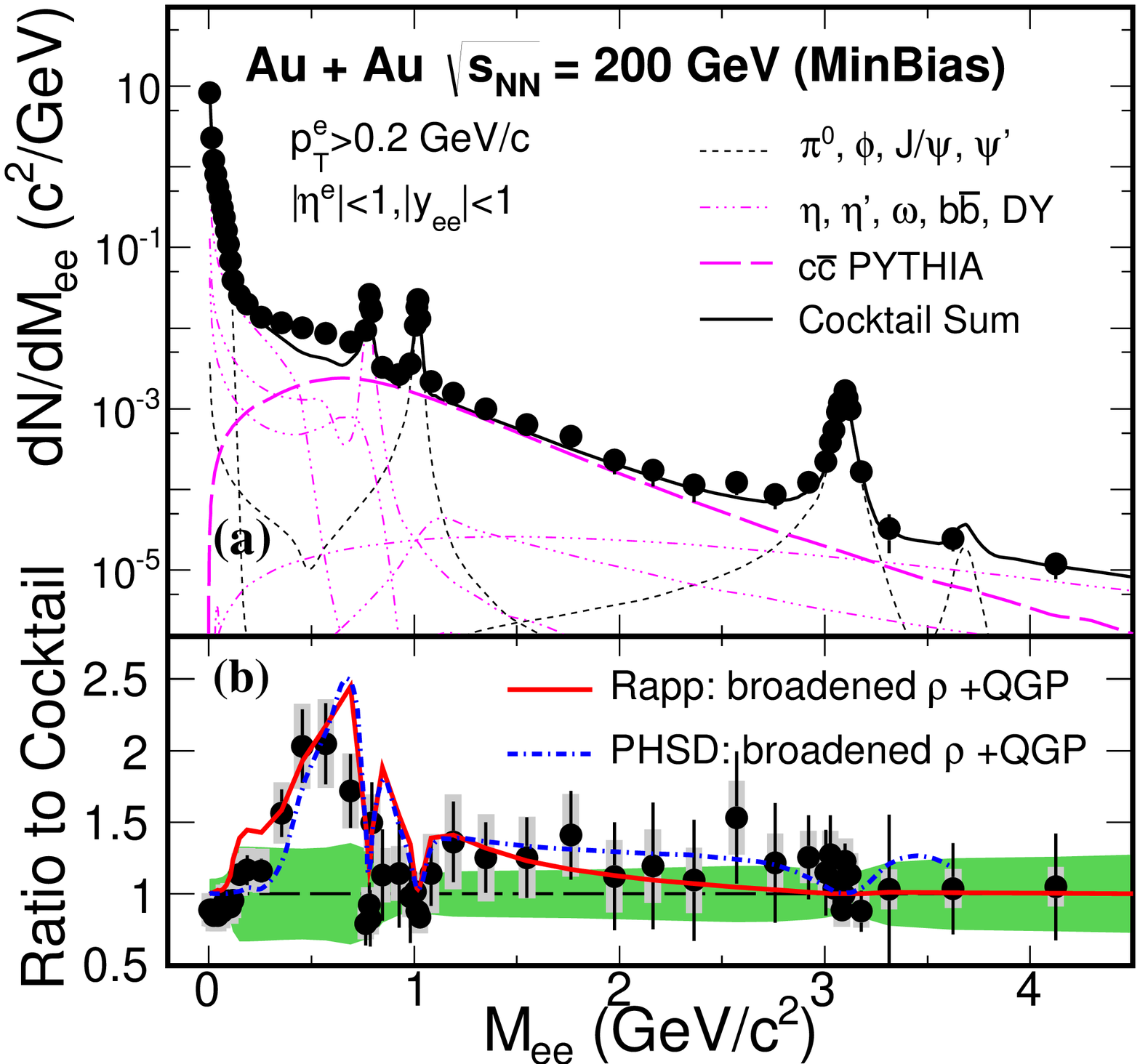}
\includegraphics[width=0.45\textwidth]{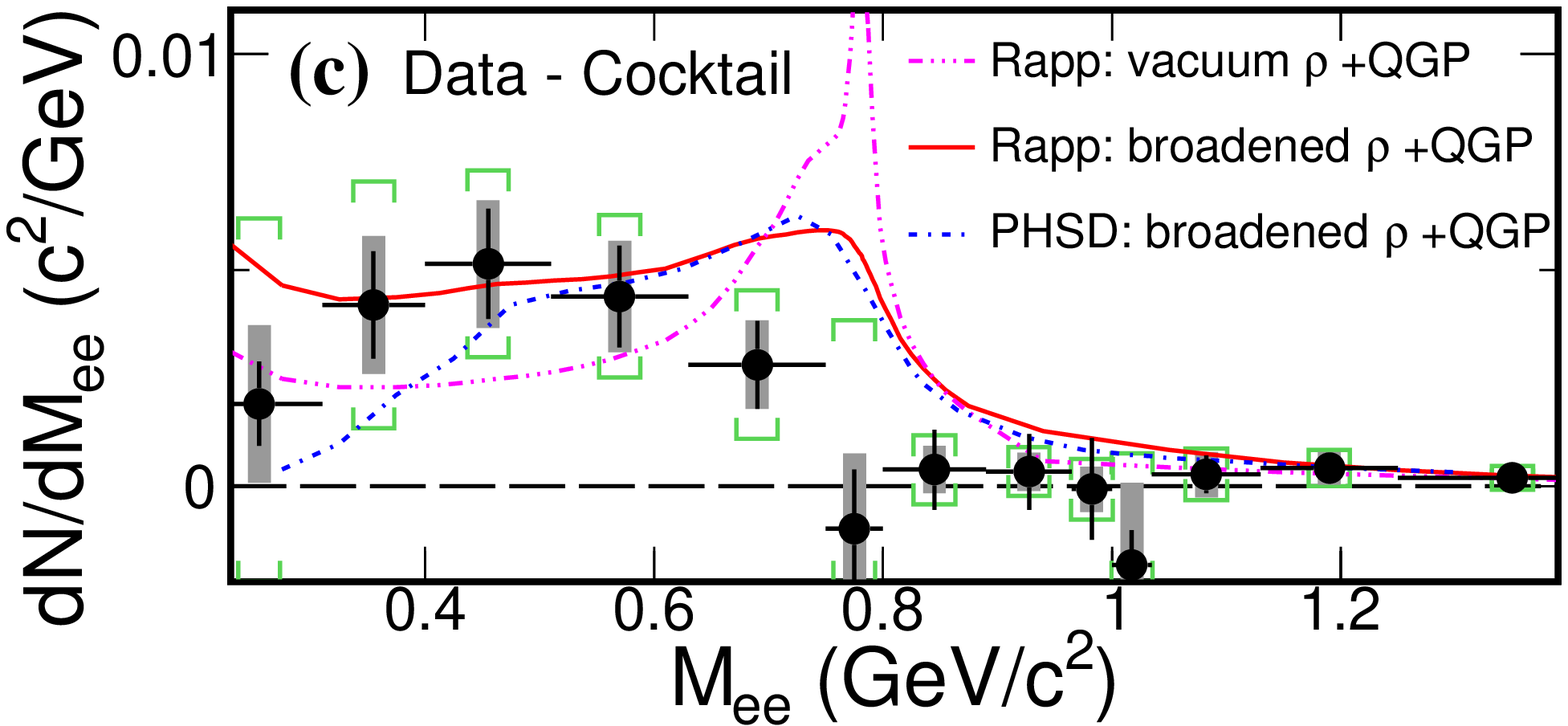}
\caption{(Color online) (a) $e^+e^-$ invariant mass spectrum from \sNN = 200 GeV Au+Au minimum bias (0-80\%) collisions compared to a hadronic cocktail simulation.
The vertical bars on data points depict the statistical uncertainties, while the systematic uncertainties are shown as grey boxes (smaller than the marker).
(b) Ratios to cocktail for data and model calculations~\cite{RappPriv,PHSDEE}. Green bands
depict systematic uncertainties on the cocktail.
(c) Mass spectrum of the excess (data minus cocktail) in the low-mass region compared to model calculations. Green brackets depict the total systematic uncertainties including those from cocktails. Systematic errors are highly correlated across all data points.
}
\label{invMass}
\end{figure}

In Fig.~\ref{invMass} (a), a comparison is shown between the hadronic cocktail simulations and the efficiency corrected dielectron yield in 200\ GeV minimum bias Au+Au collisions, in the STAR acceptance range of $p_{\rm T}^e>$0.2 GeV/$c$, $|\eta^e|<$1 and $|y_{ee}|<$1. The hadronic cocktail simulations exclude contributions from the $\rho$ meson to avoid double counting when compared to models. The ratios of our measured data to the cocktail are shown in panel (b) of Fig.~\ref{invMass}. Panel (c) in Fig.~\ref{invMass} shows an expanded view of the excess mass region with the cocktail subtracted.
An enhancement of 1.77$\pm$0.11(stat.)$\pm$0.24(sys.)$\pm$0.33(cocktail) is observed when compared to the hadronic cocktail without the $\rho$ contribution in the mass region of 0.30-0.76\,GeV/$c^2$. This enhancement factor, determined within the STAR acceptance, is significantly lower than what has been reported by PHENIX ~\cite{PHENIX}.
We have compared the STAR and PHENIX cocktail simulations and applied PHENIX azimuthal acceptance. We found that neither differences in the acceptance nor the cocktail simulations can explain the difference in the enhancement factor measured by the two experiments.

Also included in Fig.~\ref{invMass} (b) and (c) are two theoretical model calculations within the STAR acceptance: Model I by Rapp {\it et al.} is an effective many-body calculation~\cite{RappThermalEE,Rapp4SPS,RappPriv};
Model II by Linnyk {\it et al.} is a microscopic transport model - Parton-Hadron String Dynamics (PHSD)~\cite{PHSD,PHSDEE,PHSD4SPS}. Both models have successfully described the dimuon enhancement observed by NA60 with a broadened $\rho$ spectral function due to in-medium hadronic interactions. The models, however, failed to reproduce the dielectron enhancement reported by PHENIX~\cite{PHENIX,PHSDEE}. Compared to our data in the mass region below 1\,GeV/$c^2$, both models describe the observed dielectron excess reasonably well within uncertainties.
Other theoretical model calculations can also reproduce the dielectron excess at low-mass in our measurement~\cite{HJXu,Vujanovic}.
Our measurements disfavor a pure vacuum $\rho$ spectrum for the excess dielectrons ($\chi^{2}$/NDF= 26/8, where NDF is the number of degrees of freedom, in 0.3-1\,GeV/$c^{2}$).

\begin{figure}[h]
\centering
\includegraphics[width=0.45\textwidth] {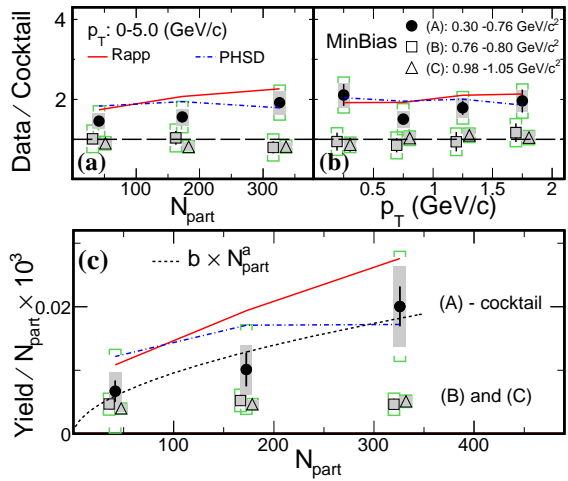}
\caption[]{(Color online) Integrated dielectron yields in the mass regions of 0.30-0.76\,($\rho$-like), 0.76-0.80\,($\omega$-like) and 0.98-1.05\,($\phi$-like) \GeVcsq compared to hadronic cocktails within the STAR acceptance as a function of centrality (a) and dielectron \pT (b). Panel (c) shows the yields scaled by $N_{\rm part}$ for the $\rho$-like region with cocktail subtracted, the $\omega$-like and $\phi$-like regions without subtraction as a function of $N_{\rm part}$. Systematic uncertainties from data are shown as grey boxes, and green brackets depict the total systematic uncertainties including those from cocktails. For clarity, the $\omega/\phi$-like data points are slightly displaced horizontally.}
\label{enhanceCent}
\end{figure}

We integrated the dielectron yields in three mass regions: 0.30-0.76\,($\rho$-like),
0.76-0.80\,($\omega$-like) and 0.98-1.05\,($\phi$-like)\,GeV/$c^2$, and present the centrality and \pT dependence of the ratios of data to cocktail within the STAR acceptance in Fig.~\ref{enhanceCent} (a) and (b).
The cocktail calculation can reproduce the dielectron yields in the $\omega$-like and the $\phi$-like regions.
The ratios to cocktail in the $\rho$-like region show a weak dependence on the number of participating nucleons ($N_{\rm part}$) and $p_{\rm T}$. Both models show excesses comparable to the data in the centrality and $p_{\rm T}$ regions investigated.
Panel (c) shows the integrated yields scaled by $N_{\rm part}$ for the $\rho$-like with cocktail subtracted, the $\omega/\phi$-like without subtraction as a function of $N_{\rm part}$. The $\omega/\phi$-like yields show a $N_{\rm part}$ scaling. The dashed curve depicts a power-law fit ($\propto N_{\rm part}^a$) to the $N_{\rm part}$ scaled $\rho$-like dielectron excess with cocktail subtracted, and the fit result shows $a=0.54\pm0.18$ (stat.+uncorrelated sys.), indicating dielectrons in the $\rho$-like region are sensitive to the QCD medium dynamics, as expected from $\rho$ medium modifications in theoretical calculations~\cite{RappPriv,RhoOmega}.

\begin{figure}[h]
\centering
\includegraphics[width=0.45\textwidth] {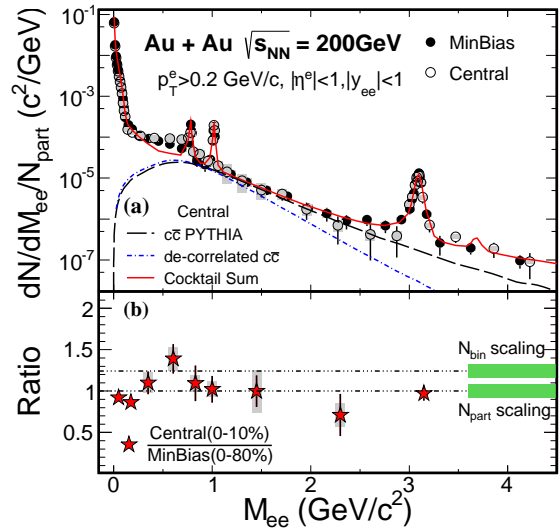}
\caption[]{(Color Online) (a)Dielectron mass spectra scaled with $N_\mathrm{part}$ from minimum bias (0-80\%) and central (0-10\%) collisions. The solid line represents the hadronic cocktail for central collisions. (b) The ratio of $N_\mathrm{part}$ scaled dielectron yields between the central and minimum bias collisions. The grey boxes show the systematic uncertainties on the data.}
\label{dielectronRcp}
\end{figure}

Figure~\ref{dielectronRcp} shows a comparison of the invariant mass spectra between 0-80\% minimum bias and 0-10\% most central Au+Au collisions. Both spectra were scaled by $N_\mathrm{part}$ in panel (a), and the ratios of the two scaled spectra are presented in panel (b).
Horizontal bands on the right side depict the $N_\mathrm{part}$ and $N_\mathrm{bin}$ scaling.
We note that:
(i) The dielectron production starts with the $N_{\rm part}$ scaling in the $\pi^{0}$ and $\eta$ dominant region and then rises towards the $N_{\rm bin}$ scaling at $\sim$0.7\ GeV/$c^2$.
This can be explained by the hadronic medium $\rho$ contribution, which is expected to increase faster than $N_\mathrm{part}$~\cite{RappPriv,RhoOmega}, and the contribution from correlated charm which, if not modified, should follow the $N_{\rm bin}$ dependence. Possible charm de-correlation has negligible impact in this mass region.
(ii) In the mass region of 1-3\ GeV/$c^2$, the ratio between the central and minimum bias spectra shows a moderate deviation from the $N_{\rm bin}$ scaling (1.8$\sigma$ deviation for the data point at 1.8-2.8 GeV/$c^2$).
We have used two extreme scenarios to model the charm decay dielectron pairs: The dashed line in Fig.~\ref{dielectronRcp} (a) depicts the PYTHIA calculation with charm correlations preserved; the dot-dashed line assumes a fully randomized azimuthal correlation between charmed hadron pairs and the \pT suppression factor on the single electron spectrum from RHIC measurements is also included~\cite{NPERaa}.
The difference in the mass region 1-3 GeV/$c^2$, if confirmed with better precision, would constrain the magnitude of the de-correlating effect on charm pairs while traversing the QCD medium and/or possible other dielectron sources (e.g., QGP thermal radiation) in this mass region from central Au+Au collisions.

In summary, we present STAR measurements of dielectron production in Au+Au collisions at \sNN = 200 GeV.
The dielectron yields in the $\omega$ and $\phi$ mass regions are well described by the hadronic cocktail model while yields at higher mass, 1-3\,GeV/$c^2$, can be understood as mostly from decay leptons of charm pairs. In the 0.30-0.76\,GeV/$c^2$ region, however, there exists a clear excess over the hadronic cocktail that cannot be explained by a pure vacuum $\rho$. This enhancement is significantly lower than what has been reported by PHENIX.  Compared to the yields in the $\omega$ and $\phi$ regions, the excess yields in the $\rho$ region exhibit stronger growth with more central collisions.
Theoretical model calculations that include a broadened $\rho$ spectral function from interactions with the hadronic medium can describe the STAR measured dielectron excess at the low-mass region.

We thank the RHIC Operations Group and RCF at BNL, the NERSC Center at LBNL, the KISTI Center in Korea,
and the Open Science Grid consortium for providing resources and support. This work was supported in part
by the Offices of NP and HEP within the U.S.DOE Office of Science, the U.S. NSF, CNRS/IN2P3, FAPESP CNPq of
Brazil,the Ministry of Education and Science of the Russian Federation, the NNSFC, the MoST of China (973 Program
No. 2014CB845400), CAS, the MoE of China, the Korean Research Foundation, GA and MSMT of the Czech
Republic, FIAS of Germany, DAE, DST, and CSIR of India, the National Science Centre of Poland, National
Research Foundation (Grant No. NRF-2012004024), the Ministry of Science,Education and Sports of the Republic of
Croatia, and RosAtom of Russia.



\end{document}